# IMAGE CLASSIFIERS FOR NETWORK INTRUSIONS

David A. Noever and Samantha E. Miller Noever

PeopleTec, Inc., Huntsville, Alabama, USA
david.noever@peopletec.com

## ABSTRACT

This research recasts the network attack dataset from UNSW-NB15 as an intrusion detection problem in image space. Using one-hot-encodings, the resulting grayscale thumbnails provide a quarter-million examples for deep learning algorithms. Applying the MobileNetV2's convolutional neural network architecture, the work demonstrates a 97% accuracy in distinguishing normal and attack traffic. Further class refinements to 9 individual attack families (exploits, worms, shellcodes) show an overall 56% accuracy. Using feature importance rank, a random forest solution on subsets show the most important source-destination factors and the least important ones as mainly obscure protocols. The dataset is available on Kaggle.

## 1. INTRODUCTION

This work updates the UNSW-NB15 attack dataset [1-4] and extends the popular intrusions detection system (IDS) originally inspired by the KDD-99/DARPA challenge [5-7]. The details of the UNSW-NB15 dataset are published in a series of previous papers [1-4] which described the raw network packet captures, generated features on labeled attacks, and scored statistical methods for identifying each attack family. As illustrated in Figure 1, the current approach aims to map scaled numerical features to images, a method likened to traditional spectrogram methods. These fingerprinting techniques have proven useful when image-based neural networks have solved similar but challenging time-dependent [8] or audio [9] problems. We test the capabilities for mapping tabular features to build fast image classifiers. One advantage of this hierarchical method arises from the unique power of transfer learning to high accuracy, even when the underlying patterns prove difficult for humans to understand or classify. The dataset is available on Kaggle [10].

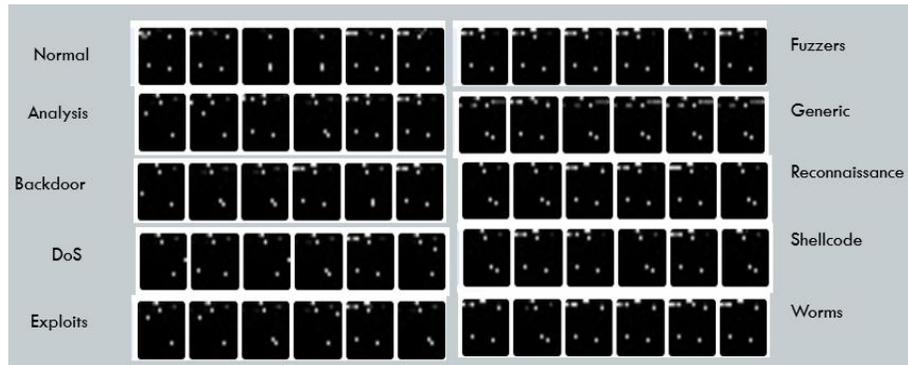

Figure 1. Nine attack types and one normal traffic dataset. We map the tabular features to grayscale thumbnails.

## 2. METHODS

The use of convolutional neural networks for network attack classification depends on first converting all tabular feature sets into thumbnail images. We, therefore, recast the UNSW-NB15 tabular set of features as scaled image thumbnails [11] to solve for 9 attack types. This version of the dataset renders the corresponding UNSW attack set as 256-pixel grayscale images (16 x16). We employ one-hot-encoding [12] for the categorical inputs and rescale all numerical inputs as grayscale pixel values (0-255) between the training set's minimum and maximum values. The baseline UNSW-NB15 dataset [1-4] yields 194 values and the images are right padded with all black (255) values for any unused pixels (62) identically for all attack labels. This padding assists deep learning approaches [13] which have a stride length in powers of 2. The column labels are also included in the train and test sets as tabular formats (comma-separated value files) to compare image-based classification methods to more statistical approaches like decision trees, random forest, and support vector machines. The expectation is that all the legacy algorithms of both deep learning and statistical machine learning may assist in the new task after mapped to images of feature sets. This approach shares many characteristics with the traditional MNIST dataset [14-20] and thus can build quickly on those findings for algorithmic comparisons. Several image-based problems to solve include simply binary classifiers for attack vs. normal traffic. Like MNIST digits [14], there are 10 categories shown (0=normal; 1-9 various attacks). As shown in Figure 2, the original 42 network features expand to 194 when one-hot-encoded [12]. This process converts all categorical data (services, protocols, and states) into individual columns with their presence marked by 1 and absence by 0. For instance, protocol_http becomes pixel value 255 at the appropriate grayscale image location (row=3, column = 13) if the attack used hypertext transfer protocol. Conversely, the same pixel maps to 0 if the protocol was not used. Each row of the UNSW thus renders 256 features (of which 194 follow directly from the tabular set).

|    | 1 | 2 | 3 | 4 | 5 | 6 | 7 | 8 |
|----|---|---|---|---|---|---|---|---|
| 1  | dur | spkts | dpkts | sbytes | dbytes | rate | sttl | dttl |
| 2  | swin | stcpb | dtcpb | dwin | tcprtt | synack | ackdat | smean |
| 3  | ct_dst_src_ltm | is_ftp_login | ct_ftp_cmd | ct_flw_http_mthd | ct_src_ltm | ct_srv_dst | is_sm_ips_ports | service_- |
| 4  | service_smtp | service_snmp | service_ssh | service_ssl | protocol_3pc | protocol_a | n | protocol_aes-sp3-d |
| 5  | protocol_cbt | protocol_cftp | protocol_chaos | protocol_compaq | protocol_cphb | protocol_cpnx | protocol_crtp | protocol_crudp |
| 6  | protocol_etherip | protocol_fc | protocol_fire | protocol_ggp | protocol_gmtp | protocol_gre | protocol_hmp | protocol_i-nlsp |
| 7  | protocol_il | protocol_ip | protocol_ipcomp | protocol_ipcv | protocol_ipip | protocol_iplt | protocol_ipnip | protocol_ippc |
| 8  | protocol_iso-ip | protocol_iso-tp4 | protocol_kryptolan | protocol_l2tp | protocol_larp | protocol_leaf-1 | protocol_leaf-2 | protocol_merit-inp |
| 9  | protocol_nsfnet-igp | protocol_nvp | protocol_ospf | protocol_pgm | protocol_pim | protocol_pipe | protocol_pnni | protocol_pri-enc |
| 10 | protocol_sat-expak | protocol_sat-mon | protocol_sccopmce | protocol_scps | protocol_sctp | protocol_sdrp | protocol_secure-v | protocol_sep |
| 11 | protocol_stp | protocol_sun-nd | protocol_swipe | protocol_tcf | protocol_tcp | protocol_tlsp | protocol_tp++ | protocol_trunk-1 |
| 12 | protocol_vrrp | protocol_wb-expak | protocol_wb-mon | protocol_wsn | protocol_xnet | protocol_xns-idp | protocol_xtp | protocol_zero |
| 13 | pad | pad | pad | pad | pad | pad | pad | pad |
| 14 | pad | pad | pad | pad | pad | pad | pad | pad |
| 15 | pad | pad | pad | pad | pad | pad | pad | pad |
| 16 | pad | pad | pad | pad | pad | pad | pad | pad |

|    | 9 | 10 | 11 | 12 | 13 | 14 | 15 | 16 |
|----|---|----|----|----|----|----|----|----|
| 1  | sload | dload | sloss | dloss | sinpkt | dinpkt | sjit | djit |
| 2  | dmean | trans_depth | response_body_le | ct_srv_src | ct_state_ttl | ct_dst_ltm | ct_src_dport_ltm | ct_dst_sport_ltm |
| 3  | service_dhcp | service_dns | service_ftp | service_ftp-data | service_http | service_irc | service_pop3 | service_radius |
| 4  | protocol_any | protocol_argus | protocol_aris | protocol_arp | protocol_ax.25 | protocol_bbn-rcc | protocol_bna | protocol_br-sat-mon |
| 5  | protocol_dcn | protocol_ddp | protocol_ddx | protocol_dgp | protocol_egp | protocol_eigrp | protocol_emcon | protocol_encap |
| 6  | protocol_iatp | protocol_ib | protocol_idpr | protocol_idpr-cm | protocol_idrp | protocol_ifmp | protocol_igmp | protocol_igp |
| 7  | protocol_ipv6 | protocol_ipv6-fra | protocol_ipv6-no | protocol_ipv6-opt | protocol_ipv6-rou | protocol_ipx-n-ip | protocol_irtp | protocol_isis |
| 8  | protocol_mfe-nsp | protocol_mhrp | protocol_micp | protocol_mobile | protocol_mtp | protocol_mux | protocol_narp | protocol_netblt |
| 9  | protocol_prm | protocol_ptp | protocol_pup | protocol_pvp | protocol_qnx | protocol_rdp | protocol_rsvp | protocol_rvd |
| 10 | protocol_skip | protocol_sm | protocol_smp | protocol_snp | protocol_sprite-r | protocol_sps | protocol_srp | protocol_st2 |
| 11 | protocol_trunk-2 | protocol_ttp | protocol_udp | protocol_unas | protocol_uti | protocol_vines | protocol_visa | protocol_vmtp |
| 12 | state_ACC | state_CLO | state_CON | state_FIN | state_INT | state_REQ | state_RST | pad |
| 13 | pad | pad | pad | pad | pad | pad | pad | pad |
| 14 | pad | pad | pad | pad | pad | pad | pad | pad |
| 15 | pad | pad | pad | pad | pad | pad | pad | pad |
| 16 | pad | pad | pad | pad | pad | pad | pad | pad |

Figure 2. Layout template for one-hot-encoded images

We leave unchanged the train/test split of the original UNSW-NB15 dataset at a 1:2 ratio [1-4]. The detailed counts for each class are shown in Figure 3. It is worth noting that the UNSW-NB15 dataset updates and statistically rebalances some of the KDD99 counts based on their analysis of duplicates and potential data leakage between training and test sites. The ratio of 1:2 for training and test presents a challenging amount of previously unseen data when an algorithm gets scored or deployed. In total, we created almost a quarter-million images as 256-pixel thumbnails using ImageMagick [11]. The largest training class (normal traffic: 37,000) outnumbers the smallest attack class (worms: 44) by nearly 1000:1 as a ratio of cases.

| Training | |
|---|---|
| Attack | 45332 |
| Normal | 37000 |
| | 82332 |
| Testing | |
| Attack | 119341 |
| Normal | 56000 |
| | 175341 |

| Family | Count |
|---|---|
| Analysis | 677 |
| Backdoor | 583 |
| DoS | 4089 |
| Exploits | 11132 |
| Fuzzers | 6062 |
| Generic | 18871 |
| Normal | 37000 |
| Reconnaissance | 3496 |
| Shellcode | 378 |
| Worms | 44 |
| Grand Total | 82332 |

Figure 3 Training and testing count per attack family count

To explore whether transfer learning from a convolutional neural network can identify network attacks, we tested the small (2 Mb) MobileNetV2 model [13] as pre-trained, then introduced both the binary and multi-class problems. The binary classifier determines whether a given image pattern represents normal or attack traffic. The multi-class problem identifies one of the 10 possible families (9 attacks in Figure 3 vs. normal). The multi-class example shares an analogous data setup to the traditional MNIST handwriting dataset [14] and thus may benefit from the various state-of-the-art approaches developed to handle those 10 classes. We solve both the binary and multi-class cases with a standard set of hyper-parameters (epochs:50, batch size:16, learning rate: 0.001). Slower learning rates disrupt the pre-trained layers of the neural network and preserve some of its beneficial weights for feature extraction in the images. We also explore the effects of smaller dataset size (<10,000 training examples vs. the full 250k) [22].

To explore the feature importance for detecting attacks, we applied a random forest algorithm (Figure 4) to the binary classifier [23]. The rank order for the top 14 contributors is shown using the Gini Index (or impurity) which effectively gauges the factors contributing to a decision split between normal and attack [23]. The highest contributors include 1) the Source to destination time to live value (sttl); 2) Number for each state (dependent protocol, e.g. ACC, CLO, CON) according to a specific range of values for source/destination time to live (ct_state_ttl); and 3) Number of connections of the same source and the destination address in 100 connections according to the last time (ct_dst_src_ltm). Not shown in Figure 4 are the least important which somewhat surprisingly include most of the one-hot-encoded protocol features that are more exotic than ordinary TCP (e.g. zero, XTP, XMN.IDP, WSN, etc.). In addition to providing a future path to reduce the intrusion detector's dimensionality, this feature importance rank defines what cannot be safely ignored in attack datasets like UNSW-NB15 [1-4].

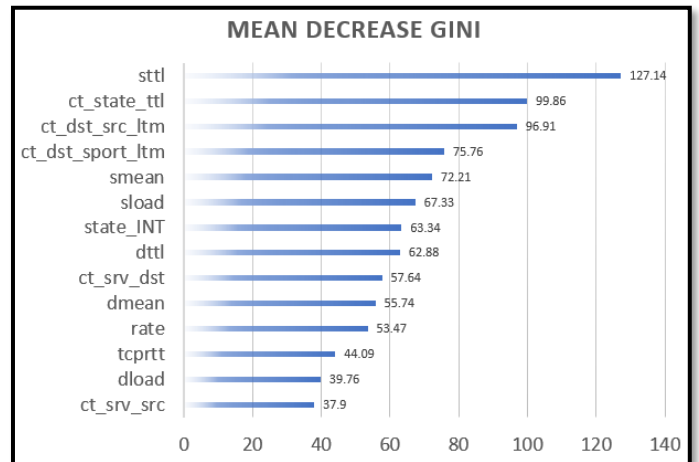

Figure 4 Feature information (GINI) Contribution to Attack Detection

## 3. RESULTS

As shown in Figures 4-6, the results for transfer learning with a deep convolutional neural network (MobileNetV2 [13]) demonstrate that the image-based binary classifier achieves greater than 97% accuracy in identifying whether an attack occurs (Figure 5).

| Family | Accuracy | Test |
|---|---|---|
| Normal | 98% | 500 |
| Attack | 97% | 542 |

Figure 5 Binary classifier results

In Figure 6, the specific identification of an attack family averages 54% accuracy between the 10 classes, with large deviations between the best (normal and generic traffic: 96+%) and the worst (worms and analysis <19%). Figure 6 shows the error matrix of which attack images confuse the neural network (worms misclassed as exploits). It is worth noting that the majority class (normal) loses no performance as more attack classes get added from the binary to the multi-classification example.

| Family | Accuracy | Test | Accuracy | Test | Family | Accuracy | Test | Accuracy | Test |
|---|---|---|---|---|---|---|---|---|---|
| Analysis | 19% | 102 | 21% | 39 | Generic | 96% | 226 | | 0 |
| Backdoor | 69% | 88 | 44% | 41 | Normal | 98% | 473 | 67% | 36 |
| DoS | 34% | 161 | 54% | 41 | Reconn | 63% | 176 | 5% | 44 |
| Exploits | 66% | 264 | 44% | 50 | Shellcode | 21% | 57 | 79% | 42 |
| Fuzzers | 70% | 204 | 82% | 39 | Worms | 0% | 7 | 14% | 7 |

Figure 6 Multi-class results

One contribution to this variance is the relative sparsity of UNSW-NB15 examples for the lower performing classes. To test this hypothesis, we performed the same experiment on a smaller subset (<4000) of images with a hold-out test and validation set that represents 20% of the training set (as opposed to 200% in the original UNSW-NB15 split). By better balancing, the dataset, undetectable worms, and other lesser represented classes could be detected in Figure 6 (second column). Mapping attacks to images shows the dependence of accuracy on both class size and imbalance [23].

One interesting outcome of using these image-based detection maps is their portability to small hardware appliances. The small MobileNetV2 architecture [13] is tailored to run on edge devices [24], such as mobile phones. Simple network detectors thus render a complex matrix of packet features into a rapid classifier capable of running in near real-time imagery (e.g. 30 frames -or attacks- per second). The reduction of the model to use tflite (Tensorflow) as a set of stored weights represents a standard model [25] for deploying deep learning to edge devices.

| | Normal | Analysis | Backdoor | DoS | Exploits | Fuzzers | Generic | Reconn | Shellcode | Worms |
|---|---|---|---|---|---|---|---|---|---|---|
| Normal | 98% | 0% | 0% | 0% | 1% | 1% | 0% | 0% | 0% | 0% |
| Analysis | 0% | 19% | 51% | 10% | 19% | 2% | 0% | 0% | 0% | 0% |
| Backdoor | 0% | 14% | 69% | 5% | 9% | 1% | 0% | 2% | 0% | 0% |
| DoS | 0% | 2% | 0% | 34% | 59% | 2% | 0% | 2% | 1% | 0% |
| Exploits | 1% | 0% | 1% | 19% | 66% | 3% | 0% | 9% | 0% | 0% |
| Fuzzers | 0% | 0% | 3% | 8% | 9% | 70% | 0% | 8% | 0% | 0% |
| Generic | 0% | 0% | 0% | 0% | 2% | 0% | 96% | 0% | 0% | 0% |
| Reconn | 0% | 1% | 1% | 14% | 17% | 3% | 0% | 63% | 2% | 0% |
| Shellcode | 0% | 0% | 0% | 0% | 0% | 5% | 0% | 74% | 21% | 0% |
| Worms | 0% | 0% | 0% | 0% | 86% | 14% | 0% | 0% | 0% | 0% |

Figure 7. Confusion matrix by class

## 4. DISCUSSION AND CONCLUSIONS

The results demonstrate a viable path of converting tabular feature data to image thumbnails, then applying convolutional neural networks to classify attack families. One potential shortcoming in our approach is any dependencies on the parametric ordering in the table format. For instance, CNNs tend to highlight close neighbors as being related in the image [26], yet there is no obvious relationship in the generated images between protocols, services, or states that justify making them into a particular attack fingerprint. One could address this flaw quantitatively by shuffling the order and determining the change of accuracy (if any). Future work should compare alternative statistical methods borrowed from the extensive machine learning literature devoted to the MNIST (and its derivative [14-21]) dataset of handwriting recognition. One can anticipate that like MNIST solutions, there exist high accuracy decision trees (like extreme gradient boosted trees – XGBoost [27]) that generate both accuracy and inference speeds comparable to the deep learning approach here. Further work could also use the image dataset [28] to design new attacks (and defenses) based on the techniques of generative adversarial networks (GANs [28]). The dataset is available on Kaggle [10].

**ACKNOWLEDGMENTS**

The author would like to thank the PeopleTec Technical Fellows program for encouragement and project assistance.